\pdfoutput=1
\documentclass{article}

\usepackage[final]{nips_2018}
\usepackage{graphicx} 

\usepackage[utf8]{inputenc} 
\usepackage[T1]{fontenc}    
\usepackage{hyperref}       
\usepackage{url}            
\usepackage{booktabs}       
\usepackage{amsfonts}       
\usepackage{nicefrac}       
\usepackage{microtype}      
\usepackage{hyperref}
\usepackage{url}
\usepackage{graphicx}
\usepackage{natbib}
\usepackage{xcolor}

\title{Avoiding a Tragedy of the Commons \\ in the Peer Review Process}

\author{
  D. Sculley, Jasper Snoek, Alex Wiltschko \\
  \texttt{\{dsculley, jsnoek, alexbw\}@google.com} \\
  Google AI \\
}

\begin{document}

\maketitle

\vskip-0.2in

\begin{abstract}
Peer review is the foundation of scientific publication, and the task of reviewing has
long been seen as a cornerstone of professional service.
However, the massive growth in the field of machine learning has put this community
benefit under stress, threatening both the sustainability of an effective review process and
the overall progress of the field.  In this position paper, we argue
that a {\em tragedy of the commons} outcome may be avoided by emphasizing 
the {\em professional} aspects of this service.  
In particular, we propose a rubric to hold reviewers to an objective standard for review quality.
In turn, we also propose that reviewers be given appropriate incentive.  As one possible
such incentive, we explore the idea of financial compensation on a per-review basis.
We suggest reasonable funding models and thoughts on long term effects.
\end{abstract}

\section{Introduction: the Current State of Peer Reviews in Machine Learning}
\begin{figure}
    \centering
   \label{neurips}
     \caption{{\bf Submissions Per Year for NeurIPS (left) and ICLR (right).} Both conferences have seen exponential growth in submissions per year, straining the pool of experienced reviewers.}
    \includegraphics[width=1.8in]{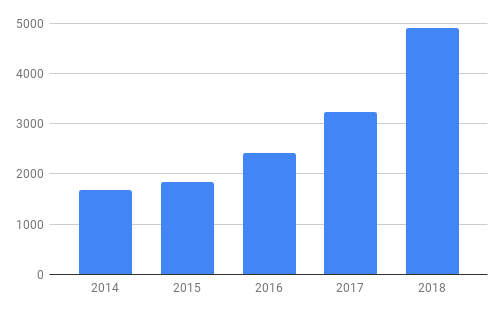} 
    \includegraphics[width=1.8in]{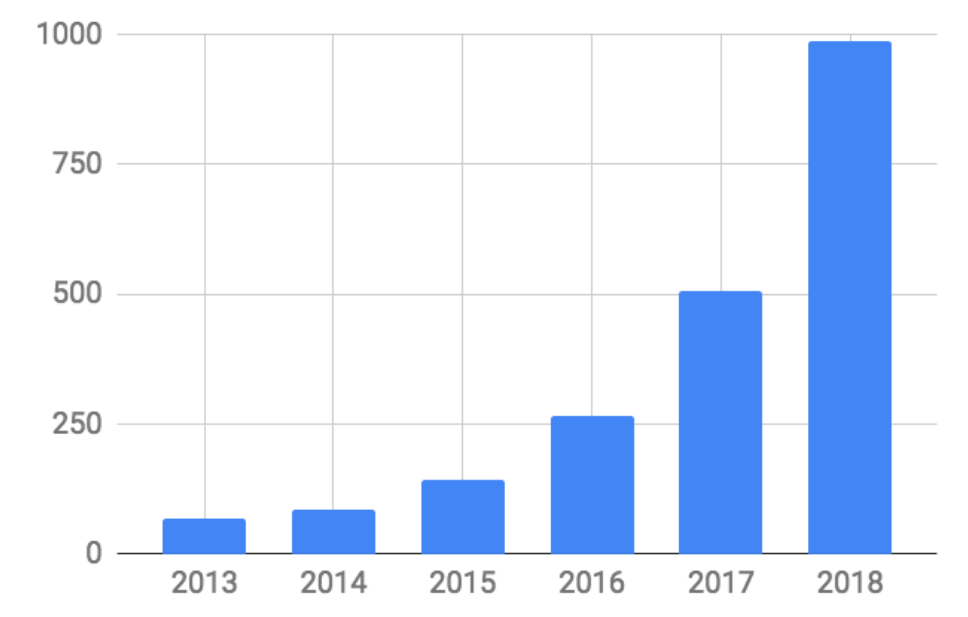} 
    \vskip-0.25in
\end{figure}

Machine learning as a field has been undergoing incredible growth in recent years in all dimensions,
including new findings and levels of participation, but perhaps most striking is the growth in conference paper submissions.  In the last year alone, the number of conference paper submissions needing review has increased by 47\% for ICML, by 50\% for NeurIPS, and by almost 100\% for ICLR.  (See Figure~\ref{neurips}.)  This exponential growth rate has undoubtedly strained the community, particularly in its capacity to perform timely and thorough peer review \citep{lipton2018troubling,sculley2018winners}.
This is alarming: peer review is the primary mechanism by which we enforce rigorous standards for our scientific community.  But the increased strain on experienced reviewers recalls the {\em tragedy of the commons}, in which a limited communal resource is laid waste from overuse.

Indeed, we may already be at a breaking point. In mid-2018 we looked at hundreds of past NeurIPS and ICLR reviews\footnote{Reviews at:
\url{https://openreview.net/group?id=ICLR.cc/2018/Conference} and \url{http://papers.nips.cc/book/advances-in-neural-information-processing-systems-30-2017}} to find high quality reviews to use as exemplars in an updated reviewer guide.\footnote{Guide at: \url{https://nips.cc/Conferences/2018/PaperInformation/ReviewerACSACGuidelines}}  We had hoped to find many reviews that evaluated \emph{quality}, \emph{clarity}, \emph{originality} and \emph{significance} in a clear and thorough way, but were surprised at the difficulty of finding individual reviews that satisfied all three. Instead we ended up using review snippets highlighting just one of of these areas at a time.  Overall, this proved to be a sobering exercise, but one that gave a realistic view of the current state of ML reviewing.

This is far from an isolated anecdote.  
The NeurIPS Experiment of 2014, conducted by Fawcett and Cortes,\footnote{See Eric Price's summary at \url{http://blog.mrtz.org/2014/12/15/the-nips-experiment.html}} found that when papers were randomly re-reviewed by independent committees of reviewers within the NeurIPS reviewing process, almost 60\% of the papers accepted by the first committee were rejected by the second.  
The dramatic increase in submissions and the corresponding reliance on less experienced reviewers is likely to have only exacerbated this effect.

We believe the the key issues here are structural.  Reviewers donate their valuable time and expertise anonymously as a service to the community with no compensation or attribution, are increasingly taxed by a rapidly increasing number of submissions, and are held to no enforced standards.  

Thus, we propose a structural solution to this issue, with two key components:
\begin{itemize}
    \item Reviewers should be held to objective standards for review quality, as determined by a rubric that can be applied consistently by area chairs and other senior committee members. 
    \item Reviewers should be incentivized and rewarded for their work appropriately, commensurate with market rates for their time and level of expertise. 
\end{itemize}
 We give an example rubric for consideration in Section 2, and Section 3 provides a detailed look at financial compensation as one possible mechanism to allow reviewers to prioritize this important work ahead of competing demands on their time.

\section{A Rubric for Evaluating Peer Reviews}
Rubrics have long been used in education~\cite{wolf2007rubric} to evaluate student performance since they can make evaluation more objective, consistent and fair. They also help students understand what is expected of them by making the evaluation criteria explicit~\cite{andrade2005} and thus can act as a guide to communicate expectations to reviewers.  We believe such a rubric, detailing specific expectations, with gradations of review quality would help evaluate incentivized reviews and improve review quality overall.  

As a starting point, we provide a seven point rubric for review quality in Figure~\ref{rubric} with a checklist for components of high quality reviews.  We believe this quick checklist could be applied with low effort by area chairs to provide a reasonably objective assessment of review quality.

\section{Incentives and Compensation}

The obvious cost of enforcing the high standards from the rubric in Figure~\ref{rubric} is one of time and added effort.  Proper incentives are important if this cost is to be borne in a sustainable way.  While other strong incentives such as free or reserved registrations are reasonable \citep{sculley2018winners}, here we examine the idea of using direct financial incentives to make clear that such an approach is indeed feasible.

Regardless of the specific mechanism, we do view this professional service as one worthy of professional compensation.  The goal is to provide sufficient reward to empower reviewers to prioritize their reviewing responsibility over other competing demands on their time, while equally empowering area chairs and other senior committee members or chairs with an appropriate mechanism to motivate compliance with high standards of reviewing.

\paragraph{What is a reasonable level of compensation?} To make this proposal concrete, we would suggest a figure of USD 1000 per paper per review as a reasonable value of compensation.
We arrive at this figure by noting that a complete review including reading, understanding, verifying, and writing should take at least half a day (4 hours) of attention from a qualified expert in the field.  Assuming a rate of USD 250 per hour as a conservative estimate for consulting work done by experts in machine learning\footnote{One estimate: \url{https://towardsdatascience.com/how-to-price-an-ai-project-f7270cb630a4}} in the current market gives a value of USD 1000 per review per paper.  As with other honoraria, some reviewers will likely decline, especially those with industrial research affiliations.

\paragraph{Would every review qualify?}  We do not propose financial compensation for its own sake, but rather as an important incentive for review excellence.  Thus, we suggest that area chairs use a rubric like that in Section 2 to assess the quality of each review on an objective scale, and that only those reviews passing a high bar for excellence be rewarded with payment.  Area chairs would be encouraged to iterate with reviewers when necessary to help reviews meet the bar.  The ability to withhold compensation would be an explicit and important dis-incentive against hasty, incomplete, late, or otherwise low quality reviews.

\paragraph{Alternate incentives.}
Financial compensation is not the only form of reasonable incentive.  We note with encouragement other forms of incentive such as reserved registrations that have been made available to the top 1000 reviewers for NeurIPS 2018, and ``top reviewer'' awards at ICML, both of which are good examples of non-monetary incentives.  However, an ideal incentive is one that rewards additional work in a smoothly increasing way.  For example, a reviewer who does a great job with 10 reviews should be rewarded more than one who does a great job with 5, and either reviewer should ideally see additional benefit if they were to do one more.  Discrete incentives such as reserved registrations introduce a step function that detracts from this effect, while financial incentives increase smoothly as reviewers contribute additional value.

\paragraph{Related compensation practices.}
It is worth noting other fields do incorporate compensation into their peer review and publication processes. For instance, in fields where publication is mainly journal-based, professional editors are employed to oversee the initial screening of submitted work, selecting individual reviewers, and adjudicating disputes in the review process itself. These editors usually have PhDs in the journal's field.

Although compensation for journal article reviewing is rare, the National Institute of Health (NIH), a major government funding agency for research in biology and medicine, pays an honorarium for peer review activities for grants \cite{nih-honorarium}. Grant review at the NIH occurs over a short and concentrated time-period, resembling the conference review process in machine learning. In-person, electronic and mail-in participation in the review process are all compensated with an honorarium. Neither the presence of paid editorial oversight nor compensated peer review have been raised as impediments to objectivity in the fields that employ them.

\begin{figure}
\caption{{\bf A 7 point rubric for evaluating review quality.} This checklist can be used by area chairs or other senior committee members.}
\begin{itemize}
\sc
\small
\item[$\square$] Review establishes the relevant expertise of the reviewer
\item[$\square$] Review demonstrates understanding of the contribution via concise summary
\item[$\square$] Review evaluates the quality of the writing and composition of the work
\item[$\square$] Review places the contribution in current context of prior publications and evaluates novelty
\item[$\square$] For each proof or formal claim, review demonstrates checked correctness or demonstrates flaw
\item[$\square$] For each empirical result, review evaluates experimental protocols, verifies baselines and evaluates statistical significance
\item[$\square$] Review evaluates actual significance of the work and interest to the community
\end{itemize}
    \vskip-0.2in
\label{rubric}
\end{figure}

\section{Funding and Budgeting}

Compensating USD 1000 per review per paper creates a material cost that must be funded, on the order of USD 15 million for a conference like NeurIPS 2018.  Here we give several alternatives by which this could be supported, along with ideas for reducing the overall figure. 

\paragraph{Differential pricing for conference registration.}
NeurIPS 2018 famously sold out within roughly 6 minutes of registration opening.  From an economic point of view, it appears that demand far exceeds supply, and the classic response would be to raise prices.  However, we do not believe that it is appropriate to make it more difficult for students or academics to register; promoting an inclusive community is critical to the long-term vitality of the field.  

Instead, we suggest that it may be reasonable to offer {\em differential pricing}, for example charging significantly more for industry-affiliated attendees than for academic researchers or current students.  To begin the conversation, imagine a world in which industry-affiliated attendees were charged a figure on the order of USD 5000.  This would provide a budget of the correct magnitude.  As beneficial side effects, it may also mitigate some of the registration crunches and provide a mechanism to further promote inclusion and diversity at registration time.

\paragraph{Co-pay with submission.}
A co-pay is a classic solution to help ensure that consumers do not abuse low cost resources.\footnote{See \url{https://en.wikipedia.org/wiki/Copayment}} In this setting, paper submission would entail a fee which would both be used both to collect funds to help offset reviewing cost, while also providing a disincentive for authors to submit work that is not yet fully ready for review.  Such methods should obviously not be used in ways that would deter participation from authors who would be overly burdened by such a cost; as suggested by \citep{sculley2018winners} we would recommend that a payment-in-kind mechanism by which authors can provide paper reviews in place of a monetary co-pay be made available as an additional alternative.

\paragraph{Explicit sponsorship.}
Industry sponsorship currently provides more than USD 1.6 million to NeurIPS to fund various events.  Given the value of machine learning research to the industrial world, this may be seen as a surprisingly low figure. We believe it may be reasonable to create sponsorship opportunities to help fund review costs, and to highlight these as the most prestigious sponsors.  Such framing may be sufficient to create a sufficient budget.  However, because of the possibility of perceived bias toward sponsoring institutions, we consider direct sponsorship a less desirable option.

\paragraph{Incentivizing efficiency.}  One benefit of providing compensation for reviews is that it concretely highlights the true value of reviewer time, and may motivate measures to ensure that this time is used as efficiently as possible.  Conference review tools are often seen as difficult to work with, but the question of whether to invest in improving them may become more clear if the consideration is a 10\% improvement in reviewer time on a multi-million dollar budget.

\paragraph{Reducing time spent on low-quality papers.} Currently, a large fraction of reviewer time is spent on providing high quality reviews for papers that are far below the acceptance threshold for a given conference.  In 2018, there were 1366 NeurIPS submissions averaging an overall score of 4 or below.  These consumed more than 4000 reviews, or the full review capacity of nearly 700 reviewers, despite the fact that these papers are essentially certain of rejection.  We suggest dispensing with the custom that all papers receive the same number of reviews.  Multi-stage triage that quickly removes papers far below the bar may reduce the reviewing burden substantially, while improving the ability of reviewers to dive deeply into the details of higher quality papers.  This is common practice in journal-centric fields like biology and medicine, where professional editors perform triage for the review process and filter out papers clearly below the bar from further review.

\section{Possible Long Term Effects}

In current practice, peer reviews are unfunded and are often distributed to exactly those people who are least economically advantaged in the current research community, such as students nearing completion of their PhDs who are currently required by tradition to provide this value for free.   We believe that compensating reviewers for their reviews will help such members of the community by providing an important additional source of funding which rewards expertise and may also help enable such researchers to take greater risks and pursue more ambitious directions.

Additionally, current practice is to involve a large number of reviewers, and cap review load (typically at six papers), in order to avoid overly taxing any individual reviewer.  This load capping makes it difficult for individual reviewers to be well calibrated to the standards of a given conference or to compare a range of similar submissions in a hot topic for a given year, and actively requires broadening the reviewer pool to less expert reviewers if all relevant experts are fully capped.  In a system in which reviewers may actually prefer to review more papers in order to increase their compensation, we may end up with reviewers deciding to take on 10 or 20 reviews.  Such reviewers may have better perspective than is currently typical, and may be more efficient on a per-review basis when reviewing many papers on a similar topic.

\section*{Acknowledgements}
We thank Samy Bengio, George Dahl, Hugo Larochelle, Jamie Morgenstern, and especially the anonymous reviewers for this paper for their helpful discussions and feedback.  We also thank 
Tara Sainath for providing the ICLR submissions-per-year data and for her permission to reproduce the ICLR chart from her slides.

Finally, after initial publication we found an earlier independent use of the phrase ``tragedy of the commons'' in conjunction with thoughts on peer review process in the field of Ecology from \citep{tragedyecology}, focusing on problems connected with authors overloading the journal review process with papers that may not be not fully ready for full review.  It is perhaps some comfort to know that machine learning is not alone as a field in facing such considerations. 

\bibliographystyle{unsrt}
\bibliography{main}

\begin{thebibliography}{1}

\bibitem{lipton2018troubling}
Zachary~C. Lipton and Jacob Steinhardt.
\newblock Troubling trends in machine learning scholarship.
\newblock In {\em International Conference on Machine Learning: The Debates},
  2018.

\bibitem{sculley2018winners}
D.~Sculley, Jasper Snoek, Alex Wiltschko, and Ali Rahimi.
\newblock Winner's curse? on pace, progress, and empirical rigor.
\newblock In {\em International Conference on Learning Representations
  Workshops}, 2018.

\bibitem{wolf2007rubric}
Kenneth Wolf and Ellen Stevens.
\newblock The role of rubrics in advancing and assessing student learning.
\newblock {\em The Journal of Effective Teaching}, 7:3--14, 2007.

\bibitem{andrade2005}
H.~Andrade and D.~Ying.
\newblock Student perspectives on rubric-referenced assessment.
\newblock {\em Practical Assessment, Research and Evaluation}, 10(3):1--11,
  2005.

\bibitem{nih-honorarium}
Non-federal peer review travel guidelines.
\newblock
  \url{https://grants.nih.gov/grants/peer/guidelines_general/NonFederalPeerReviewTravelGuidelines.pdf},
  2018.
\newblock Accessed: 2018-10-30.

\bibitem{tragedyecology}
Michael~E Hochberg, Jonathan~M Chase, Nicholas~J Gotelli, Alan Hastings, and
  Shahid Naeem.
\newblock The tragedy of the reviewer commons.
\newblock {\em Ecology Letters}, 12(1):2--4, 2009.

\end{thebibliography}

\end{document}